\documentclass[12pt]{article}
\usepackage[final]{graphics}
\usepackage{epsfig}
\usepackage{amsmath, amsthm, amssymb}
\usepackage{cite}
 \usepackage{latexsym}
\usepackage{sparticles}
 \usepackage{feynmf}
 \usepackage{hep}
\newcommand {\nc} {\newcommand}
\nc {\tstrut} {\vline  height 2.5ex depth 1.7ex width 0ex}
\nc {\ustrut} {\vline  height 2.0ex depth 0.5ex width 0ex}
\numberwithin{equation}{section}
\begin{document}
 \title{Helicity amplitudes and crossing relations\\ for antiproton proton reactions}
\author{N.~H.~Buttimore\,\, and \,\,E.~Jennings\footnote{elise@maths.tcd.ie}
\\ \emph{School of Mathematics, Trinity College, Dublin, Ireland}}
\maketitle
\abstract{
\noindent Antiproton proton annihilation 
reactions allow  unique access
to the moduli and phases of nucleon electromagnetic form factors 
in the time like region. We present the helicity amplitudes
for the unequal mass single photon reaction $p\,\bar{p}\to l^+\,l^-$ in the 
$s$ channel including the lepton mass. The relative signs of these 
amplitudes are determined using simple invariance properties.
Helicity amplitudes for the annihilation reaction  
$p\,\bar{p} \to N\,\bar{N}$ are also given, where $N$ is any spinor 
particle with structure.
Crossing relations between the $e\,p \to e\,p$ scattering and 
the $p\,\bar{p}\to l^+\,l^-$ annihilation channels are discussed
and the crossing matrix for the helicity amplitudes is given.
This matrix may be used to verify known expressions for the space 
like helicity amplitudes due to one photon exchange.
}

\section{Introduction}

	Electromagnetic form factors contain information on the
internal dynamics and structure of the nucleon.
Form factors provide a description of baryons as composites and are
 related to the Fourier transforms of charge and magnetisation
 density distributions of the nucleon.
 These fundamental functions of momentum transfer indicate the hadron's
internal structure,
allowing us to probe the quark core as
 well as providing information on the parton model.

	At present there is considerable interest in this area mainly due to
the surprising result that, for example,
the proton time like form
factor is found  to be twice as large as in the space like region
\cite{Andreotti:2003}
and  that evaluations of  the electromagnetic
form factor ratio $G_E/G_M$ show a deviation
from Rosenbluth scaling.
Recent measurements of the electric proton form factor at large values
of momentum transfer squared, $q^2$, show discrepancies
in the results depending on whether a recoil polarisation method
was used or the spin averaged differential cross section in
electron proton scattering \cite{Jones:2000,Punjabi:2005wq}.
It has been suggested that these results
may  be resolved if we consider the  two photon exchange
 mechanism taking place in the reaction as well as performing more
experiments in the time like region in order to access the phases of
these form factors.
Recent discussions on this subject \cite{Ambrosino:2006gk, Buttimore:2006mq}
 as well as extraction of proton form factors 
in the time like region \cite{Bianconi:2006wg}
have reviewed and highlighted the
need for further study and experimentation in this area.

	The Polarised Antiproton Experiment (PAX)
collaboration \cite{Barone:2005pu} aims to
shed new light on nucleon electromagnetic
form factors by accessing  single and double spin observables.
They plan to do this by using polarised antiprotons, produced by spin
filtering with an internal polarised gas target.
It is through these experiments that they hope to measure
the moduli and the relative phase of the
time like electric and magnetic form factors of the proton.
The extensive spin programme at RHIC (BNL) also aims to provide new information 
on nucleon electromagnetic form factors \cite{RHIC1,RHIC2}.

	While there are several papers which discuss and
present equations for the moduli and phases of the proton form
factors in the time like region,
\cite{Tomasi-Gustafsson:2005kc},
\cite{Dubnickova:1992ii} and \cite{Buttimore:2006mq},
there are advantages
to considering helicity transition amplitudes for
two body annhilation reactions \cite{Jacob}.
Helicity amplitudes are particularly useful in analysing a
scattering process as conservation of angular momentum
in the centre of mass frame 
and spin dependent observables
can be expressed simply.
Also symmetry principles
such as parity conservation and time reversal invariance
can be stated directly in terms of helicity amplitudes
which allows us to easily determine the number of
independent amplitudes for a given process.

	It is well known that in the space like regime the Dirac
and Pauli form
factors are real; however in the time like region they are complex
on the real axis above the threshold $4 m_{\pi}^2$,
where $m_{\pi}$ is the pion mass.
These two regimes are connected by dispersion relations
which permit analytic continuation of the amplitudes 
from the time like
to the space like region.
It is these analytic properties which give rise to a crossing
 matrix \cite{Trueman:1964}
between the helicity amplitudes for the $t$
channel reaction $l^-\,p\to l^-\,p$ and the $s$ channel
reaction $p\,\bar{p}\to l^+\,l^-$.
This crossing matrix is presented in section~5.

	The paper is organised as follows:
in section~2 we outline the calculation and results for the
five independent helicity amplitudes for the $s$ channel reaction
$p\,\bar{p} \to l^+\,l^-$.
It is shown that the squared moduli of the amplitudes sum to the spin
averaged differential cross section for the same process,
confirming that helicity amplitudes contribute incoherently
to the unpolarised cross section.
Section 3 employs the symmetry properties of charge conjugation, 
parity and time reversal invariance in order to
determine the relative signs of these amplitudes.

	In section 4 we present the helicity amplitudes for the
 annihilation reaction $N\,\bar{N} \to p\, \bar{p}$ in terms
of four form factors.
The final section reviews the crossing relations between the
scattering, $t$, and the annihilation, $s$, channels and 
discusses the Wigner rotations needed to transform 
from one channel to another.
We give  the crossing matrix $W$ which connects  the
proton lepton scattering helicity amplitudes
to the amplitudes for proton antiproton annihilation giving 
rise to a lepton antilepton pair.
This matrix, $W$, is a $( 6 \times 6 )$ matrix whose entries depend on the
Mandelstam variables
$s$, $t$ and on $m$ and $M$ the masses of the lepton and proton
respectively. The crossing matrix $W$ reduces to a previously published result
when $m=M$ \cite{Spearman}.

\section{Helicity amplitudes for $ \l^-\,l^+ \to \bar{p}\,p$}
 \vspace{2ex}
 \begin{figure}[!htb]
 \begin{center}
 \begin{tabular}{cccccccccccccccc}
\begin{fmffile}{fig1}    
%fig1.mf will be created for this feynman diagram
\fmfframe(1,7)(1,7){    %Sets dimension of Diagram
 \begin{fmfgraph*}(110,62)
 %Sets size of Diagram
\fmfleft{i1,i2}
%Sets there to be 2 sources
 \fmfright{o1,o2}
%Sets there to be 2  outputs
\fmflabel{$l^+(K',S_2)$}{i2}
 \fmflabel{$l^-(K,S_1)$}{i1}
 \fmflabel{$p(P,S_3)$}{o2}
 %Labels one of the left sources
 \fmflabel{$\bar{p}(P',S_4)$}{o1}
 %Labels one of the left sources
\fmf{fermion}{i1,v1,i2} %Connects the sources with a vertex.
\fmf{fermion}{o1,v2,o2} %Connects the outputs with a vertex.
 \fmf{photon,label=$\gamma(q)$}{v1,v2}
\fmfdot{v1}
 \fmfblob{.16w}{v2}
 %labels the conneting line.
 \end{fmfgraph*}
 }
\end{fmffile}
 \end{tabular}
 \\
\caption{One photon exchange for 
$l^-\,+\,l^+ \to \bar{p}\,+\,p$ in 
the $s$ channel.} 
\label{fey1}
 \end{center}
 \end{figure}
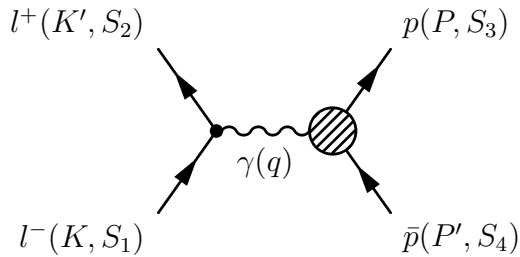
\vspace{1ex}

	In calculating the helicity amplitudes for
lepton antilepton annihilation leading to a proton antiproton
pair, figure \ref{fey1}, or the time reversed 
reaction we can firstly find
the differential cross sections for the reaction 
involving pure helicity states.
Taking the square root of this cross section yields
the absolute value of the helicity amplitude.
 We discuss the relative signs of these amplitudes in section 3.
The formulae used for the differential cross section
in terms of the invariant amplitude ${\cal M}$ is given in appendix A.

	When all four spinors relate to polarised fermions the
invariant amplitude for lepton antilepton to
proton antiproton due to one photon exchange is
 \begin{eqnarray}
 \cal{M} & = &
i\,e^2\,\bar{v}(K', S_2)\,\gamma_{\mu}\,u(K, S_1)\,\,\frac{1}{s}\,
\,\bar{u}(P, S_3)\left[\,\gamma^{\mu}\,F_1(s) + 
\frac{i}{2\,M}\,\sigma^{\mu \nu}\,q_{\nu}\,F_2(s)\,
\right]v(P', S_4)\nonumber
\\
 \end{eqnarray}
where  $M$ is the mass of the proton and 
$\sigma^{\mu\nu} = \frac{1}{2} \,i\,[\gamma^{\mu},\gamma^{\nu}]$.
 The four momenta of the incoming lepton and antilepton
 are $K^{\mu}$ and $K'\,^{\mu}$ respectively and $P^{\nu}$
 and $P'\,^{\nu}$ are the four momenta of the
 outgoing proton and antiproton.
These momenta as well as the spin four vectors, 
$S_{i}, i \in \{1,2,3,4\}$,
 are given in the centre of mass
frame for the time reversed reaction in appendix B.
Here $s = q^2 = q_{\mu}q^{\mu}$ is the square of the
invariant 4-momentum in the s channel.
So that
$q^2$ is positive in the time like region
we adopt the metric $q^2 = q^2_0 - \textbf{q}^2$.
This is given in terms of the final proton and antiproton
momenta by
\begin{eqnarray}
 q^{\nu} &=& P^{\nu} + P'\,^{\nu}\,.
\end{eqnarray}
The Dirac and Pauli form factors, $F_1$ and $F_2$,
 are analytic functions of $q^2$.
 They take real values in the space like region $q^2 < 0$
due to the hermiticity of the Hamiltonian.
The form factors are complex in the time like
 region on the real $q^2$ axis above threshold.
 \noindent Using Gordon decomposition this amplitude becomes
 \begin{eqnarray}
 \cal{M} & = &
i\,e^2\,\bar{v}(K',S_2)\,\gamma_{\mu}\,u(K, S_1)\,
 \frac{1}{s}\,
\\&&\qquad
\times\, \bar{u}(P, S_3)\,\left[\,\gamma^{\mu}\,G_M(s)
+ \frac{1}{2\,M}\,\left(\,P' - P\,\right)^{\mu}\,F_2(s)\right]\,
v(P', S_4)\,.\nonumber
\end{eqnarray}
The Sachs electric and magnetic form factors $G_E$ and $G_M$ are
 a linear combination of $F_1$ and $F_2$
and are defined in appendix A.

	Using the Mathematica package \lq\lq {\tt{Tracer.m}}\rq\rq 
$\,$ we can evaluate the proton and lepton current.
The proton current requires us to evaluate the following trace:
\begin{eqnarray}
\mbox{Tr}(\cancel{P} + M)\left(
 1 + \gamma_5 \cancel{S_3}\right)\,
 \left[\gamma^{\nu}G_M + R^{\nu}F_2\right]
 (\cancel{P'} - M)\left(1 + \gamma_5 \cancel{S_4}\right)
 \left[\gamma^{\mu}G^*_M + R^{\mu}F^*_2\right]\,
\end{eqnarray}
where $R_{\mu} = (P_{\mu}\,'- P_{\mu})/2\,M$ and $*$ denotes 
complex conjugation. 
 Here as we are not summing over proton polarisations
 we have $u(P,S_3)\,\bar{u}(P,S_3) = (\cancel{P} - M)/2 $.
The lepton current involves a slightly simpler trace,
\begin{eqnarray}
 \mbox{Tr}\left(\cancel{K} + m\right)\left(
 1 + \gamma_5 \cancel{S_1}\right)\gamma_{\mu}
\left(\cancel{K'} - m\right)\left(1 + \gamma_5 \cancel{S_2}\right)
 \gamma_{\nu}
\end{eqnarray}
where $m$ denotes the lepton mass. Once each of these traces is calculated and then
contracted together we arrive at the
differential cross section for the process.

	We shall briefly outline the notation
 used in the following sections.
The  helicity amplitudes
 describing the $s$ channel reaction
 \begin{eqnarray}
 l^- + l^+ \to \bar{p} + p
\end{eqnarray}
are written as
 \begin{eqnarray}
H\,(\lambda_{\bar{p}}\,\lambda_{p}\,;\,\lambda_{l^-}\, \lambda_{l^+})
\end{eqnarray}
 where $\lambda_{i}$ is the helicity of particle $i$,
 which is either $+$ or $-$ depending on which
helicity amplitude we are considering.
	We find the following five independent
cross sections for this process:
 \begin{eqnarray}
 \label{5indep}
 \frac{d\sigma}{d\Omega}\,(+\,+\,;\,+\,+) &=&
\frac{d\sigma}{d\Omega}\,(+\,+\,;\,-\,-) =
 \frac{d\sigma}{d\Omega}\,(-\,-\,;\,-\,-) =
 \frac{d\sigma}{d\Omega}\,(-\,-\,;\,+\,+)\nonumber
 \\
 \frac{d\sigma}{d\Omega}\,(+\,+\,;\,+\,-) &=&
 \frac{d\sigma}{d\Omega}\,(+\,+\,;\,-\,+) =
 \frac{d\sigma}{d\Omega}\,(-\,-\,;\,-\,+) =
 \frac{d\sigma}{d\Omega}\,(-\,-\,;\,+\,-)\nonumber
 \\
 \frac{d\sigma}{d\Omega}\,(+\,-\,;\,-\,-) &=&
\frac{d\sigma}{d\Omega}\,(+\,-\,;\,+\,+) =
\frac{d\sigma}{d\Omega}\,(-\,+\,;\,-\,-) =
\frac{d\sigma}{d\Omega}\,(-\,+\,;\,+\,+)\nonumber
 \\
 \frac{d\sigma}{d\Omega}\,(+\,-\,;\,+\,-) &=&
 \frac{d\sigma}{d\Omega}\,(-\,+\,;\,-\,+)\nonumber
 \\
 \frac{d\sigma}{d\Omega}\,(+\,-\,;\,-\,+) &=&
 \frac{d\sigma}{d\Omega}\,(-\,+\,;\,+\,-)\,.
 \end{eqnarray}
Taking the square root of the cross section
 gives us the absolute value of the single photon 
helicity amplitude in each case.
 The expression for the centre of mass scattering angle,
 $\theta$, depends on both the lepton and proton mass
and is given in appendix B.
\begin{eqnarray}
 \label{hel1}
 \frac{1}{\alpha}\,\mid H(+\,+\,;\,+\,+)\mid &=&
 \frac{2\,m\,M}{s}\,\cos\theta\,\mid G_E \mid \nonumber
\\\nonumber
 \\\nonumber
\frac{1}{\alpha}\,\mid H\,(+\,-\,;\,+\,-)\mid &=&
\frac{1}{2}\,\left( 1 + \cos\theta \right)\,
\mid G_M\mid\nonumber
 \\\nonumber
\\
 \frac{1}{\alpha}\,\mid H\,(+\,-\,;\,-\,+)\mid &=&
\frac{1}{2}\,\left( 1 - \cos\theta \right)\,
\mid G_M\mid
\\\nonumber
 \\\nonumber
\frac{1}{\alpha}\,\mid H(+\,+\,;\,+\,-)\mid &=&
\frac{M}{\sqrt{s}}\,\sin\theta \mid G_E\mid \nonumber
 \\\nonumber
 \\\nonumber
 \frac{1}{\alpha}\,\mid H\,(+\,-\,;\,-\,-)\mid &=&
 \frac{m}{\sqrt{s}}\,\sin\theta\,\mid G_M\mid
 \end{eqnarray}
 In the above equation we can see that
 the helicity amplitude $H(+\,-\,;\,+\,-)$
 vanishes at $\theta = 180^\circ$.
 This is what we would expect since for $\theta = 180^\circ$
 the total angular momentum of the final state
is opposite to that of the initial state.

	We may check that
the squares of the helicity amplitudes, Eq.~(\ref{hel1}),
sum to the spin averaged differential cross section
 for this reaction.
The spin averaged differential cross section
 for $l^-\,l^+ \to \bar{p} \, p$ is \cite{Buttimore:2006mq}
 \begin{eqnarray}
 \label{unpol}
 \frac{d\sigma}{d\Omega}
 & =
 &\alpha^2 \,\beta\,\frac{1}{s^3\,\left(s - 4\,M^2\right)}\,
 \Bigg\{
 \,\frac{s^2}{2}\,\left(\,s - 4\,M^2 \right)
 \,|G_{M}|^2 \,
 -\,4\,s\,m^2\,M^2\,\left(\,|G_{M}|^2 - |G_{E}|^2\,\right)
 \nonumber \\[1ex]
  &&\qquad
 +\,\left[\,\left(t\,-\,m^2 - M^2 \right)^2
 + s\,t \right]\,\left(\,s\,|G_{M}|^2 - 4\,M^2|G_{E}|^2 \right)\,
 \Bigg\}\,.
\end{eqnarray}
The variable $\beta$ is a flux factor defined in appendix A.
The spin averaged cross section can also be
written in terms of the centre of mass
 scattering angle, $\theta$, as
 \begin{eqnarray}
 \label{unpol2}
 \frac{4\,s}{\alpha^2\,\beta}\,\frac{d\sigma}{d\Omega} &=&
\left(1 + \cos^2\theta\right)\,|G_M|^2 +
 \frac{4\,M^2}{s}\,\sin^2\theta\,|G_E|^2 \nonumber
 \nonumber \\[1ex]
&&\qquad
 \frac{4\,m^2}{s}\,\sin^2\theta\,|G_M|^2 +
\frac{16\,m^2\,M^2}{s^2}\,\cos^2\theta\,|G_E|^2\,. 
 \end{eqnarray}

	This unpolarised cross section for spin 1/2
particles is related to the sum of the helicity
 amplitudes listed above as
 \begin{eqnarray}
 s\,\frac{d\sigma}{d\Omega} &=&
 \frac{1}{4}\,\sum_{\lambda}\,
|\,H\,(\lambda_{\bar{p}}\,\lambda_{p}\,;\,\lambda_{l^-}\,
 \lambda_{l^+})|^2\,.
\end{eqnarray}
The sum in the above equation is over all sixteen
 helicity amplitudes.
Time reversal invariance reduces this number to eight;
 using parity invariance lowers this again to six.
 As the reaction involves identical particles
charge conjugation invariance may be used to reduce
 the number of independent helicity amplitudes to five.

	Using the five independent amplitudes that
we found in Eq.~(\ref{hel1}) above this becomes
 \begin{eqnarray}
 \label{sum}
 s\,\frac{d\sigma}{d\Omega}_{{\scriptsize\mbox{unpol}}} &=&
|H\,(+\,+\,;\,+\,+)|^2 +
|H\,(+\,+\,;\,+\,-)|^2 +
|H\,(+\,-\,;\,-\,-)|^2 \nonumber
\\[1ex]
 && \qquad
 +\,\frac{1}{2}\, |H\,(+\,-\,;\,+\,-)|^2 +
\,\frac{1}{2}\,|H\,(+\,-\,;\,-\,+)|^2\,.
 \end{eqnarray}
 Looking at the formula for the spin averaged cross
 section given in Eq.~(\ref{unpol2}) we can see it is
 the sum of the helicity amplitudes as given in
 Eq.~(\ref{hel1}) according to Eq.~(\ref{sum}).

 \section{Relative signs of the helicity amplitudes}
 
	From the previous section we obtained the helicity amplitudes
by taking the square root of the
differential cross section.
The relative signs of these amplitudes is still to be determined.
These signs can be found using both parity and
 charge conjugation relations \cite{Leader}.
We shall firstly consider parity and use the
 symmetry properties of the $H_{\{\lambda\}}$
for reactions of the type
\begin{eqnarray*}
 A + B \to C +D
 \end{eqnarray*}
where $\lambda_A$, $\lambda_B$, $\lambda_C$
and $\lambda_D$ are the helicities of the particles
in the above reaction and $S_A$, $S_B$, $S_C$ and $S_D$
 are the spins of the particles.
 Let $\eta_j$ be the intrinsic parity of particle j.
Then we have the condition
\begin{eqnarray}
H_{-\lambda_C\,-\lambda_D\,;\,-\lambda_A\,-\lambda_B}\,(\theta) &=&
 \eta\,\left(-1\right)^{\lambda - \mu}\,
 H_{\lambda_C\,\lambda_D\,;\,\lambda_A\,\lambda_B}\,(\theta)
 \end{eqnarray}
 where
\begin{eqnarray*}
 \eta &=& \frac{\eta_C\,\eta_D}{\eta_A\,\eta_B}\,
\left(-1\right)^{S_A + S_B - S_C - S_D}
 \\
 \\
 \lambda &=& \lambda_A - \lambda_B
 \\
 \\
 \mu &=& \lambda_C - \lambda_D\,.
\end{eqnarray*}
 Then using charge conjugation for the reaction
 \begin{eqnarray*}
 A + \bar{A} \to D + \bar{D}
 \end{eqnarray*}
 we have
 \begin{eqnarray}
 H_{d\,\bar{d}\,;\,a\,\bar{a}}\,(\theta) &=&
 \left(1\right)^{\lambda - \mu}\, H_{\bar{d}\,d\,;\,
\bar{a}\,a}\,(\theta)\,.
 \end{eqnarray}
We may use these two discrete symmetry properties
to determine some of the relative signs of the helicity amplitudes.

	In \cite{peskin}
 the reaction $e^-\,e^+ \to \mu^-\,\mu^+$ is discussed,
 and the overall sign
of an amplitude is presented there.
%is determined from an elementary calculation.
 We may use this to infer the following signs for
 $\phi_3$ and $\phi_4$ 
 \begin{eqnarray}
 H(+\,-\,;\,+\,-)& =& H(-\,+\,;\,-\,+)\,=
\,-\,\frac{1}{2}\left(1 + \cos\theta\right)\, G_M
 \\
H(+\,-\,;\,-\,+)& =&  H(-\,+\,;\,+\,-)\,=
\, -\,\frac{1}{2}\left(1 - \cos \theta\right)\, G_M
 .
 \end{eqnarray}
The other helicity amplitudes in  \cite{peskin}  are zero as the mass
of the lepton is neglected there. 
% 
%We therefore must use invariance properties to
% access the signs of the other helicity amplitudes.
%
%
%
%	Using all of the above information
%we can then write the helicity amplitudes for the unequal mass
% reaction $p\,\bar{p} \to l^+\,l^-$
% %
% \begin{eqnarray}
% \label{signs}
% H(+\,+\,;\,+\,+)& =&H(+\,+\,;\,-\,-)\, =\,
%H(-\,-\,;\,-\,-)\, =\,H(-\,-\,;\,+\,+)\,=
%\,-\,\frac{2\,m\,M}{s}\,\cos\theta\,G_E \nonumber
% \\ \nonumber
% \\\nonumber
% H(+\,-\,;\,+\,-)& =& H(-\,+\,;\,-\,+)\,=\,
% -\,\frac{1}{2}\left(1 + \cos \theta\right)\, G_M
%\nonumber
% \\ \nonumber
%\\\nonumber
%H(+\,-\,;\,-\,+)& =& H(-\,+\,;\,+\,-)\,=
%\, -\,\frac{1}{2}\left(1 - \cos \theta\right)\, G_M
%\nonumber
%\\ \nonumber
% \\\nonumber
% H(+\,+\,;\,+\,-)& =& -\, H(+\,+\,;\,-\,+)\, =
% \,-\, H(-\,-\,;\,-\,+)\, =\, H(-\,-\,;\,+\,-)\,=
% \,-\,\frac{M}{\sqrt{s}}\,\sin\theta\,G_E\nonumber
% \\ \nonumber
%\\\nonumber
% H(+\,-\,;\,-\,-)& =& -\,  H(-\,+\,;\,-\,-)\, =
% \, -\,  H(-\,+\,;\,+\,+)\, =\,  H(+\,-\,;\,+\,+)\,=
% \,\frac{m}{\sqrt{s}}\,\sin\theta\,G_M\,.\nonumber
% \\
% \end{eqnarray}
%
The five independent amplitudes suggest an alternative
 labelling of helicity amplitudes in the $s$ channel
compared with the $t$ channel.
 In what follows in this paper we have assigned the labels
 $\phi_{1-6}$ to the following amplitudes
 in the annihilation reaction  $p\,\bar{p} \to l^+\,l^-$
 \begin{eqnarray}
 \phi^{(s)}_1 &=&  \,H(+\,+\,;\,+\,+) =
 \, \phantom{-}\, H(-\,-\,;\,-\,-)\nonumber
 \\\nonumber
\phi^{(s)}_2 &=& \,H(+\,+\,;\,-\,-)\, =\,
 \phantom{-}\,H(-\,-\,;\,+\,+)\nonumber
 \\\nonumber
 \phi^{(s)}_3  &=& \, H(+\,-\,;\,+\,-) =
 \phantom{-}\, H(-\,+\,;\,-\,+)\nonumber
 \\
 \phi^{(s)}_4  &=&  \, H(+\,-\,;\,-\,+) =
 \phantom{-}\, H(-\,+\,;\,+\,-)
 \\\nonumber
 \phi^{(s)}_5 &=&   H(+\,+\,;\,+\,-) =
\,- H(+\,+\,;\,-\,+)\, =\,-\, H(-\,-\,;\,-\,+)\, =
\, H(-\,-\,;\,+\,-)\nonumber
 \\\nonumber
 \phi^{(s)}_6 &=& H(+\,-\,;\,-\,-) =
-\,  H(-\,+\,;\,-\,-)\, =\, -\,  H(-\,+\,;\,+\,+)\, =
 \,  H(+\,-\,;\,+\,+)\,.\nonumber
 \end{eqnarray}
Here $\phi^{(s)}_1 = \phi^{(s)}_2$,
 which is not the case in the $t$ channel.
 We find that the definition of the amplitudes $\phi_3$ and $\phi_4$
is the same in both channels.
This re-labelling of some of the helicity amplitudes in the case
 of $\phi_5$ and $\phi_6$ may be explained by the fact
that the helicity does not change sign in the analytic continuation.
For example the amplitude $\phi^{(t)}_{5} = H\,(+\,-\,;\,+\,+)$ in the 
$t$ channel becomes $\phi^{(s)}_{5} = H\,(+\,+\,;\,-\,+)$ in the
$s$ channel.
This fact was first observed in \cite{Trueman:1964}.

 \section{Helicity amplitudes for $N\,\bar{N} \to \bar{p}\,p$}
 %So we consider the following s channel reaction.
 \vspace{1ex}
 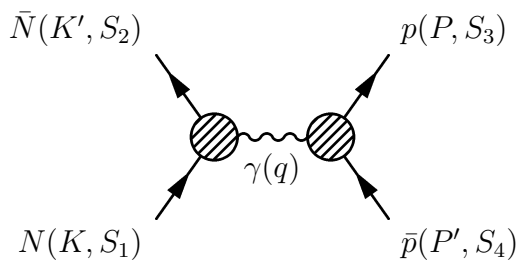
\begin{figure}[!htb]
 \begin{center}
 \begin{tabular}{cccccccccccccccc}
 \begin{fmffile}{fig2}    
%fig2.mf will be created for this feynman digram
\fmfframe(1,7)(1,7){    %Sets dimension of Diagram
\begin{fmfgraph*}(110,62)
%Sets size of Diagram
 \fmfleft{i1,i2}
 %Sets there to be 2 sources
 \fmfright{o1,o2}
 %Sets there to be 2  outputs
 \fmflabel{$N(K,S_1)$}{i1}
 %Labels one of the left sources
 \fmflabel{$\bar{N}(K',S_2)$}{i2}
 %Labels one of the left sources
 \fmflabel{$\bar{p}(P',S_4)$}{o1}
 %Labels one of the right outputs
 \fmflabel{$p(P,S_3)$}{o2}
 %Labels one of the right outputs
 \fmf{fermion}{i1,v1,i2} %Connects the sources with a vertex.
\fmf{fermion}{o1,v2,o2} %Connects the outputs with a vertex.
 \fmf{photon,label=$\gamma(q)$}{v1,v2}
 \fmfblob{.16w}{v1}
 \fmfblob{.16w}{v2}
 %labels the conneting line.
 \end{fmfgraph*}
 }
 \end{fmffile}
 \end{tabular}
 \\
\caption{One photon exchange for $N\,+\bar{N} \to \bar{p}
\,+\,p $ in  the $s$ channel.} \label{fey2}
 \end{center}
 \end{figure}

	 The next $s$ channel reaction we consider
is the $N\,+\bar{N} \to \bar{p}\,+\,p$ annihilation,
the time reverse of $\bar{p}\,+\,p \to N\,+\bar{N}$, 
figure \ref{fey2}.
 We can use the same momentum and
spin variables  as in Eq.~(\ref{mom}) and Eq.~(\ref{spin})
Using an alternative method to that used in section 2 it is possible 
to calculate the helicity amplitudes using the explicit form of the spinors.
After Gordon decomposition
the incoming nucleon current and outgoing proton current appear as
\begin{eqnarray}
u(K,S_1)\,[\,\gamma_{\mu}\,g_M + r_{\mu}\,f_2\,]\,
\bar{v}(K',S_2)\,
\bar{u}(P,S_3)\,[\,\gamma^{\mu}\,G_M + R^{\mu}\,F_2]\,v(P',S_4)
 \end{eqnarray}
 where $f_1$ and $f_2$ are the form factors of the
 incoming structured particle $N$ of mass $m$  and $g_M = f_1 + f_2$.
In the above expression $F_2$ and $G_M = F_1 + F_2$
refer to the proton form factors.
The variables $R^{\nu} = (P\,'\,^{\nu} - P^{\nu})/2\,M$
 and $r_{\mu} = (K\,'_{\mu} - K_{\mu})/2\,m$.

	Using the Dirac-Pauli representation the explicit form for the 
incoming spinors are
\begin{eqnarray}
u(K)\,=\,
\left(
\displaystyle
\begin{array}{ll}
E + m \,
\\[1ex]
\phantom{-}2\,k\,\lambda\,
\end{array}
\right)\,\Psi_{\lambda}
\,,\qquad \qquad
v(K')\,=\,
 \left(
 \displaystyle
 \begin{array}{ll}
-2\,k\,\lambda
\\[1ex]
E + m
\end{array}
\right)\,\Psi_{\lambda}\,
\end{eqnarray}
where $k$ is the magnitude of the centre of mass three momentum of the incoming particles. 
The matrix $\Psi_{\lambda}$ depends on the helicity, 
$\lambda$, and is given by  
\begin{eqnarray}
\Psi_{+}\,=\,
\left(
 \displaystyle
\begin{array}{ll}
1
\\
0
\end{array}
\right)\, ,\qquad  \qquad
\Psi_{-}\,=\,
\left(
\displaystyle
\begin{array}{ll}
0
 \\
1
\end{array}
 \right) \,.
\end{eqnarray}
In this representation the spinors of the final particles are
\begin{eqnarray}
v(P')\,=\,
\left(
 \displaystyle
 \begin{array}{ll}
-2 \,p\,\lambda \,
\\[1ex]
E + M
 \end{array}
 \right)\,\psi_{\lambda}
\, , \qquad \qquad
 u(P)\,=\,
 \left(
 \displaystyle
 \begin{array}{ll}
 E + M
  \\[1ex]
  \phantom{-}2\, p\, \lambda
\end{array}
\right)\,\psi_{\lambda}\,
 \end{eqnarray}
with
\begin{eqnarray}
\psi_{+}\,=\,
 \left(
 \displaystyle
\begin{array}{ll}
\cos \theta/2    
 \\[1ex]
\sin \theta/2     
\end{array}
\right) \, ,\qquad  \qquad
\psi_{-}\,=\,
 \left(
\displaystyle
\begin{array}{ll}
- \sin \theta/2
 \\[1ex]
\phantom{-}\cos \theta/2
\end{array}
 \right) 
\end{eqnarray}
where $E = \sqrt{s}/2$, $\theta $ is the centre of mass scattering angle and $M$ is the mass of the proton.
%The spin dependent differential cross sections for
%this process are
%\begin{eqnarray}
%\label{hel2}
%\frac{4\,s}{\alpha^2\,\beta}\,\frac{d\sigma}{d\Omega}(+\,+\,;\,+\,+)
%&=& \frac{16\, M^4}{s^2}\,\cos^2\theta\,|G_E|^2\,|g_E|^2 \nonumber
% \\\nonumber
%\\
%\frac{4\,s}{\alpha^2\,\beta}\,\frac{d\sigma}{d\Omega}(+\,+\,;\,+\,-)
%&=& \frac{4\,M^2}{s}\,\sin^2\theta\,|G_E|^2 \,|g_M|^2 \nonumber
%\\\nonumber
%\\
% \frac{4\,s}{\alpha^2\,\beta}\,\frac{d\sigma}{d\Omega}(+\,-\,;\,+\,-)
%&=& \left(1 + \cos \theta\right)^2\, |G_M|^2\,|g_M|^2 \nonumber
%\\ \nonumber
%\\ \nonumber
%\frac{4\,s}{\alpha^2\,\beta}\,\frac{d\sigma}{d\Omega}(+\,-\,;\,-\,+)
%&=& \left(1 - \cos \theta\right)^2\, |G_M|^2\,|g_M|^2 \nonumber
% \\ \nonumber
% \\ \nonumber
%\frac{4\,s}{\alpha^2\,\beta}\,\frac{d\sigma}{d\Omega}(+\,-\,;\,+\,+)
% &=& \frac{4\,M^2}{s} \sin^2\theta\,|G_M|^2\,|g_E|^2
% \end{eqnarray}
%where $g_M = \,f_1 +f_2$ and $g_E = f_1 + s/4\,m^2\,f_2$.

	Using this approach we find the 
following helicity amplitudes 
for the $N\,\bar{N}\to p\,\bar{p}$ process or its time reverse 
$p\,\bar{p} \to N\,\bar{N}$,
\begin{eqnarray}
\label{hel2} 
H\,(+\,+\,;\,+\,+) &=&
 -\,\frac{2\,\alpha\,m\,M}{s}\,\cos\theta\,G_E\,g_E\nonumber
 \\\nonumber
 \\\nonumber
H\,(+\,-\,;\,+\,-) &=&
-\, \frac{\alpha}{2}\,\left( 1 + \cos\theta \right)\,
 G_M\,g_M\nonumber
 \\\nonumber
 \\
H\,(+\,-\,;\,-\,+) &=&
 -\, \frac{\alpha}{2}\,\left( 1 - \cos\theta \right)\,
 G_M\,g_M
 \\\nonumber
 \\\nonumber
H\,(+\,+\,;\,+\,-) &=&
 -\,\frac{\alpha\,M}{\sqrt{s}}\,\sin\theta \,G_E \,g_M\nonumber
 \\\nonumber
\\
\,H\,(+\,-\,;\,+\,+) &=&
 \phantom{-}\,\frac{\alpha\,m}{\sqrt{s}}\,\sin\theta\,G_M \,g_E\nonumber
 \\ \nonumber
 \end{eqnarray}
where $g_M = \,f_1 +f_2$ and $g_E = f_1 + f_2\, s/4\,M^2$
for the spinor particle $N$.

	If $m=M$ in the above expressions then  these are
the helicity amplitudes for
the annihilation reaction $p \,\bar{p} \to p\, \bar{p}$  
with the familiar relation
 $\phi_5= -\phi_6$ for identical particle 
scattering. 
This relation is not true for the $s$ channel reaction
 $p \, \bar{p} \to l^+\,l^-$, Eq.~(\ref{hel1}).
The above equations are consistent with the previous
 equations given in Section 3 for $p\,\bar{p} \to l^+\,l^-$
 when the form factors $f_1 \to 1$ and $f_2 \to 0$.

	The spin averaged
 differential cross section for the process
$N \,\bar{N} \to p\, \bar{p}$
is reproduced by the helicity amplitudes that we found.
 \begin{eqnarray}
 \frac{d\sigma}{d\Omega} &=& \frac{\alpha^2\,\beta}{8\,s}\,
\bigg\{
 \frac{8\,m^2}{s}\,\left(\,\frac{4\,M^2}{s}\,
\cos^2\theta\,|G_E|^2 + \sin^2\theta\,|G_M|^2\,
 \right)\,|g_E|^2 \nonumber
\\[1ex] &&\qquad
 +
\,2\,\left(\,\frac{4\,M^2}{s}\,\sin^2\theta\,
|G_E|^2 + (1+\cos^2\theta) \,|G_M|^2
 \right)\,|g_M|^2
 \bigg\}\,.
 \end{eqnarray}
 This is the sum of the five
independent helicity amplitude found above as
 \begin{eqnarray}
  \frac{d\sigma}{d\Omega}_{{\scriptsize\mbox{unpol}}} &=&
 |H\,(+\,+\,;\,+\,+)|^2 +
 |H\,(+\,+\,;\,+\,-)|^2 +
 |H\,(+\,-\,;\,+\,+)|^2  \nonumber \\[1ex]
 && \qquad
+\,\frac{1}{2}\,|H\,(+\,-\,;\,+\,-)|^2 +
\,\frac{1}{2}\, |H\,(+\,-\,;\,-\,+)|^2.
\end{eqnarray}
 This result also confirms the relationship
between the 16 helicity amplitudes for $N\,\bar{N} \to p\, \bar{p}$
is the same as Eq.~(\ref{sum})
for $p\,\bar{p} \to l^+\,l^-$.
Using a linear combination of these helicity amplitudes according 
to \cite{Leader} it is possible to reproduce the known expression
for the single spin 
observable $A_N$ \cite{Buttimore:2006mq}. 
 \section{Crossing relations for the helicity amplitudes}
 
	We consider the process $A + B \to C + D$ ($t$ channel).
The crossed channel is then $B + \bar{D} \to C + \bar{A}$ 
($s$ channel) where the antiparticle of $A$ is denoted $\bar{A}$.
The crossing relation between the two channels
 helicity amplitudes is then given by \cite{Trueman:1964,Spearman}
\begin{eqnarray}
 \label{cross_rel}
&&H^{(t)}_{\lambda_C\,\lambda_D\,;\,\lambda_A\,\lambda_B}\,(s,t)
 =\nonumber 
\\[1ex] 
&&\!\! \sum_{\mu_{\bar{A}},\,\mu_{B},\,\mu_{C},\, \mu_{\bar{D}}}\,
 d^{\,t_{A}}_{\mu_{\bar{A}}\,\lambda_{A}}\,(\omega_A)\,
 d^{\,t_B}_{\mu_{B}\,\lambda_{B}}\,(\omega_B)\, 
 \,d^{\,t_C}_{\mu_{C}\,\lambda_{C}}\,(\omega_C)\,
 d^{\,t_D}_{\mu_{\bar{D}}\,\lambda_{D}}\,(\omega_D)\,
 H^{(s)}_{\mu_{C}\,\mu_{\bar{A}}\,;\,\mu_{\bar{D}}\,\mu_{B}}\,(s,t)\nonumber
\\
 \end{eqnarray}
 where $ H^{(t)}_{\lambda_C\,\lambda_D\,;\,\lambda_A\,\lambda_B}$
 is the helicity amplitude in the $t$ channel
 and $\lambda_{i}$ is the helicity of particle $i$.
 Likewise 
the helicity amplitude in the $s$ channel
is $ H^{(s)}_{\mu_{C}\,\mu_{\bar{A}}\,;\,\mu_{\bar{D}}\,
 \mu_{B}}$
and the helicities of the particles are given as $\mu_{i}$.

 \begin{figure}
 \center
 \includegraphics[height=5cm,width=9cm]{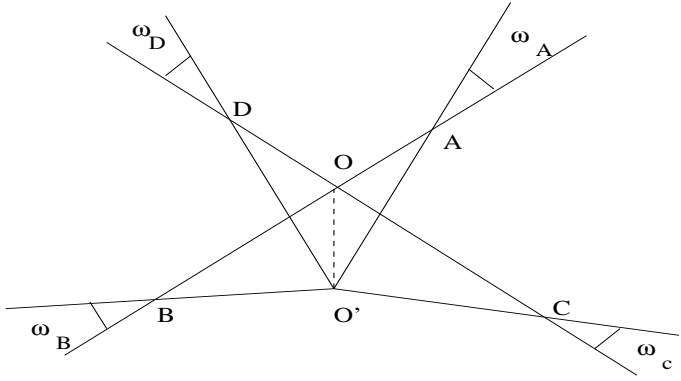}
 { \caption{
 The velocity diagram for
 the transformation in  Eq. (\ref{cross_rel}) ; $O$ is
the c.m. in the $t$ channel; $O\,'$ is the c.m.
in the $s$ channel.}}
\label{vel1}
 \end{figure}

	The rotation functions $d\,^{j}_{\mu\,\lambda}$
for spinor particles are
 \begin{eqnarray}
 j\,=\,\frac{1}{2}\,:\, &&\qquad d_{\frac{1}{2}\,
 \frac{1}{2}}\,(\beta)\,=\,\cos\frac{1}{2}\beta\,\,,
 \qquad d_{-\,\frac{1}{2}\,\frac{1}{2}}\,(\beta)\,
 =\,\sin\frac{1}{2}\beta\,\,.
\end{eqnarray}
Helicity amplitudes are generally given in a reference 
frame as they are not invariant under Lorentz transformations.
 The crossing relation for helicity amplitudes between
the $s$ and the $t$ channel is obtained by
first analytically continuing the $s$ channel
centre of mass amplitude from the $s$ to the
 $t$ physical region and then performing a Lorentz transformation
from the $s$ centre of mass frame to the $t$ centre of mass frame.
The Wigner rotations, $\omega_{i}$, for this transformation are shown
in figure 3 and
are given by
\begin{eqnarray}
\cos \omega_{A} &=&
\frac{\tstrut-\,\left(s + m^2_{A} - m^2_{B}\right)\,
\left(t + m^2_{A} - m^2_{C}\right) -
2\,m^2_{A}\,\Delta}{\tstrut {\cal S}_{A\,B}\,{\cal T}_{A\,C}}
\end{eqnarray}
where
\begin{eqnarray*}
{\cal S}^2_{i\,j} &=& [s -\left(m_i + m_j\right)^2]\,
[s -\left(m_i - m_j\right)^2]
 \\
\\
{\cal T}^2_{i\,j} &=&
 [t -\left(m_i + m_j\right)^2]\,[t -\left(m_i - m_j\right)^2]
 \\
 \\
 \Delta &=& m^2_B - m^2_{D} - m_A^2 +m^2_C\,.
 \end{eqnarray*}
Similarly the other  rotation angles are given by
\begin{eqnarray*}
 \cos \omega_{B} &=&
 \frac{\tstrut\left(s + m^2_{B} - m^2_{A}\right)\,
 \left(t + m^2_{B} - m^2_{D}\right) -
 2\,m^2_{B}\,\Delta}{\tstrut{\cal S}_{A\,B}\,{\cal T}_{B\,D}}
 \\
 \\
 \cos \omega_{C} &=&
 \frac{\tstrut\left(s + m^2_{C} - m^2_{D}\right)\,
 \left(t + m^2_{C} - m^2_{A}\right) -
 2\,m^2_{C}\,\Delta}{\tstrut{\cal S}_{C\,D}\,{\cal T}_{A\,C}}
 \\
 \\
 \cos \omega_{D} &=& \frac{\tstrut-\,
 \left(s + m^2_{D} - m^2_{C}\right)\,
 \left(t + m^2_{D} - m^2_{B}\right) -
 2\,m^2_{D}\,\Delta}{\tstrut{\cal S}_{C\,D}\,{\cal T}_{B\,D}}\,.
 \end{eqnarray*}
 Using Eq.(\ref{cross_rel}) we can the find the
 crossing matrix between the $t$ and $s$ channels.
 For lepton proton scattering $p\,l^- \to p\,l^-$
($t$ channel) there are six independent helicity
 amplitudes while for proton antiproton annihilation,
 $p\,\bar{p} \to l^+\,l^-$ ($s$ channel)
there are five independent helicity amplitudes.
We may write the crossing relation as
\begin{eqnarray}
\phi^{(t)}_{i} &=& \sum^{6}_{j=1}\, W_{i\,j}\,\phi^{(s)}_{j}
\end{eqnarray}
where $\phi^s_1 = \phi^s_2$ in the $s$ channel.

\begin{figure}
 \center
 \includegraphics[height=5cm,width=8cm]{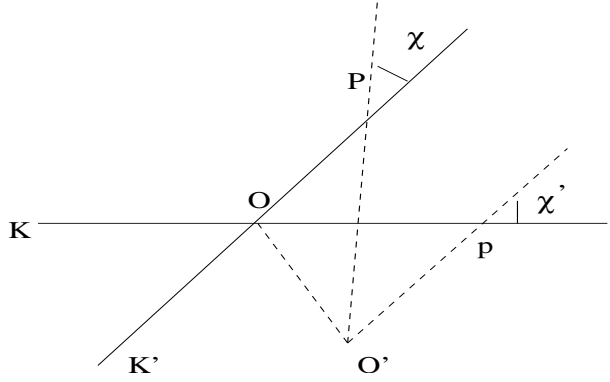}
 { \caption{
 Velocity space diagram for initial
 $t$ channel scattering; $O$ is
 the c.m. in the $t$ channel; $O\,'$ is the c.m.
 for the final configuration in the $s$ channel.}}
 \label{vel2}
 \end{figure}

	For the previous $t$ and $s$ channel reactions
 we shall use the notation of \cite{Trueman:1964}
to present the crossing matrix, $W$, as
\vspace{1ex}
 {\scriptsize {\begin{eqnarray}
 \label{M2}
W = 
 \frac{1}{2}\left(
\displaystyle
\begin{array}{llllllllllllllllllllllllllllllllllllllllll}
 \phantom{-}\sin \psi \sin \chi
 &
 \phantom{-}\sin \psi \sin \chi
 &
 1 - \cos \psi \cos \chi
 &
 1 + \cos \psi \cos \chi
 &
 -2\sin \psi \cos \chi
&
 \phantom{-}2 \cos \psi \sin \chi
 \\ [3ex]
1 - \cos \psi \cos \chi
 &
 - 1 - \cos \psi \cos \chi
&
 \phantom{-} \sin \psi \sin \chi
 &
 -  \sin \psi \sin \chi
 &
 - 2 \cos \psi \sin \chi
 &
 \phantom{-}2 \sin \psi \cos \chi
 \\ [3ex]
 \phantom{-} \sin \psi \sin \chi
 &
 \phantom{-} \sin \psi \sin \chi
 &
 - 1- \cos \psi \cos \chi
 &
 - 1 + \cos \psi \cos \chi
 &
 -2\sin \psi \cos \chi
 &
 \phantom{-}2 \cos \psi \sin \chi
 \\ [3ex]
 1+ \cos \psi \cos \chi
&
 - 1 + \cos \psi \cos \chi
 &
 - \sin \psi \sin \chi
&
 \phantom{-}\sin \psi \sin \chi
 &
 \phantom{-}2 \cos \psi \sin \chi
 &
-2  \sin \psi \cos \chi
\\ [3ex]
 - \sin \psi \cos \chi
 &
 - \sin \psi \cos \chi
 &
 -  \cos \psi \sin \chi
 &
\phantom{-}\cos \psi \sin \chi
&
 - 2  \sin \psi \sin \chi
 &
- 2  \cos \psi \cos \chi
 \\ [3ex]
 -  \cos \psi \sin \chi
&
 -  \cos \psi \sin \chi
&
 - \sin \psi \cos \chi
 &
 \phantom{-}\sin \psi \cos \chi
 &
 \phantom{-}2  \cos \psi \cos \chi
 &
 \phantom{-}2  \sin \psi \sin \chi
\end{array}\right)
\\[3ex]\nonumber
 \end{eqnarray}
}}
 where the matrix entries are given in
 terms of  Mandlestam variables as 
\begin{eqnarray}
 \cos \chi &=& -\,\cos \chi\,' \,\,=\,\,
\,\frac{\left(s + M^2 -m^2\right)\,t}
 {\tstrut \sqrt{t\,(t-4\,M^2)\,[s-\left(M+m\right)^2]\,
 [s - \left(M-m\right)^2]}}
 \\ [2ex]
 \cos \psi &=& -\,\cos \psi\,' \,\,=\,\,
 \frac{\left(s + m^2- M^2\right)\,t}
 {\tstrut \sqrt{t\,(t-4\,m^2)\,[s-\left(M+m\right)^2]\,
 [s - \left(M-m\right)^2]}}\,.
 \end{eqnarray}
Here $m$ and $M$ are the masses of the
lepton and proton respectively.
These are just the general Wigner rotations given before but for
the specfic reactions $p\,\bar{p}\to l^+\,l^-$ and
$N\,\bar{N}\to p\,\bar{p}$.
When $M=m$ in Eq.~(\ref{M2})
the matrix reduces to the already known result for
 $W$ \cite{Spearman}.
The inverse of this crossing matrix is also discussed in \cite{okorokov} as a means 
of accessing the helicity amplitudes for $p\,\bar{p}$ 
scattering using the amplitudes for the 
$p\,p$ reaction.

	The angle $\chi$ is shown in figure \ref{vel2}. 
In this diagram the
momentum of the incoming particles in
the $s$ channel is $K$ and $K\,'$ while the
outgoing momenta are $P$ and $P\,'$.
In the $t$ channel the incoming momenta are
simply $P$ and $K$ while the outgoing momenta are
$p = - P\,'$ and $k = - K\,'$. The point $O$ in figure \ref{vel2} 
represents
the centre of mass frame of $P$ and $p$ while $O\,'$ is the centre
 of mass of $P$ and $K$.

% \begin{figure}
 %\center
%\includegraphics[height=4.5cm,width=8cm]{angles.eps}
%\caption{
% Velocity space diagram for initial
%$t$ channel scattering; $O$ is
%the c.m. in the $t$ channel; $O\,'$ is the c.m.
%for the final configuration in the $s$ channel}
%\label{vel2}
% \end{figure}
%

	As the helicity amplitudes for
 proton scattering in the $t$ channel are
 known \cite{O'Brien:2006zt} we may use this matrix, Eq.~(\ref{M2})
in order to verify the signs for the
crossed channel amplitudes Eq.~(\ref{hel2}) given in previous sections.
 We have found that using the above matrix Eq.~(\ref{M2})
 and the helicity amplitudes given
 in Eq.~(\ref{hel1}) and Eq.~(\ref{hel2})
 we arrive at the correct $t$ channel amplitudes.
 We note here that using the crossing matrix and the
expressions for the helicity amplitudes in the
 $s$ channel may be used as an alternative route in accessing the
 equivalent expressions in the $t$ channel
\cite{O'Brien:2006zt, Buttimore:1978}.

\section{Conclusions}

	Proposed measurements of the electromagnetic
form factors at PAX focus on annihilation reactions in the time like 
region. Detailed knowledge of these form factors is essential in 
order to test our current understanding of pQCD asymptotics in the 
time like region and dispersion theory. Recent experiments have raised 
serious issues concerning  the electromagnetic form factors of hadrons.
It is thought that further measurements in the 
time like region will shed light on these problems 
and allow us to improve the current 
form factor models.

	Helicity transition amplitudes for proton antiproton 
electromagnetic reactions have been presented in this 
paper in the one 
photon approximation. 
These amplitudes contribute to the spin dependent observables
for the elastic reaction  
in addition to hadronic amplitudes.
We presented the helicity amplitudes for the reaction 
$p\,\bar{p}\to l^+\,l^-$ including the lepton mass. As
 expected we found five independent amplitudes for this reaction. 
These results will be useful when considering mu or tau lepton pairs
whose mass cannot be neglected.
Expressions for the unequal mass reaction $N\,\bar{N}\to p\,\bar{p}$ 
 were given in terms of four form factors. 
We found simple expressions for the five independent amplitudes 
which were found to sum to the spin averaged differential cross section for the reaction.

	The helicity amplitudes in the time like region are simply
related by analytic continuation to those in the space like region.
We discussed the rotations needed to connect these two regions
as well as presenting the  $( 6 \times 6 )$ crossing matrix between the 
$s$ and $t$ channel. This matrix together with the helicity amplitude
formulae given in this paper provide another way to compute the 
electromagnetic helicity amplitudes for the $e\,p\to e\,p$ and,
more generally, the $p\,N \to p\,N$ elastic scattering reaction.

\section{Acknowledgments}
EJ would like to thank the Irish Research Council for
Science, Engineering and Technology (IRCSET)
for a postgraduate research studentship
and Trinity College Dublin for the award of a Scholarship.
NHB is grateful to Enterprise Ireland for International
 Collaboration Programme grants that partially funded visits
 to Brookhaven National Laboratory at Upton, New York
 and to INFN at the University of Torino, Italy.
\appendix

\section{Appendix A}
 
 \noindent  The unpolarised differential cross
section for s channel annihilation of 
spin half particles in the centre of mass system is
\begin{eqnarray}
\frac{d\sigma}{d\Omega}
& = & \frac{\beta}{64\pi^2 s}\,
\frac{1}{4}\,\sum_{\mbox{\tiny spin}} |{\cal M}|^2
\end{eqnarray}
\textwidth 14cm
where $\cal{M}$ is the invariant amplitude
for the process and $\beta$ is a flux factor.
The Mandlestam variable $s$ is given
in terms of the proton and antiproton momenta by
\begin{eqnarray}
s &=& \left(P_{\mu} + P\,'_{\mu}\right)^2
\end{eqnarray}
In an annihilation reaction of two
spinor particles of mass $m_i$ producing a
pair of mass $m_f$ the flux factor $\beta$ is given by
\begin{eqnarray}
\beta& = &\left( \frac{s - 4\,m_f^2}{s - 4\,m_i^2}\right)^{1/2}.
\end{eqnarray}

\noindent The Dirac and Pauli form factors,
 $F_1$ and $F_2$, are functions of $q^2$
 and normalised at the origin such
that $F_1(0) = 1$ and $F_2(0) = \mu_p - 1$ where $\mu_p$
is the magnetic moment of the proton.
The Sachs electric and magnetic
form factors are given by \cite{Sachs:1 962}
 \begin{eqnarray}
G_E(s) & = & F_1(s) + \frac{s}{4\,M^2}\,F_2(s)
 \\ \nonumber
 \\
 G_M(s) & = & F_1(s) + F_2(s)\,.
\end{eqnarray}
By definition these form factors are equal
at threshold ($s=4\,M^2$).
In the Breit frame, $G_E$ and $G_M$
may be interpreted as the Fourier
transforms of the charge and magnetisation
 distributions, respectively.

\section{Appendix B}
  \begin{figure}[!htb]
  \begin{center}
 \begin{tabular}{cccccccccccccccc}
\begin{fmffile}{fig5}    
%5.mf will be created for this feynman diagra                m
 \fmfframe(1,7)(1,7){    %Sets dimension of Diagram
 \begin{fmfgraph*}(110,62)
  %Sets size of Diagram
 \fmfleft{i1,i2}
 %Sets there to be 2 sources
 \fmfright{o1,o2}
 %Sets there to be 2  outputs
 \fmflabel{$p(P,S_3)$}{i1}
 %Labels one of the left sources
 \fmflabel{$\bar{p}(P',S_4)$}{i2}
 %Labels one of the left sources
 \fmflabel{$l^+(K',S_2)$}{o1}
 %Labels one of the right outputs
\fmflabel{$l^-(K,S_1)$}{o2}
%Labels one of the right outputs
 \fmf{fermion}{i1,v1,i2} %Connects the sources with a vertex.
 \fmf{fermion}{o1,v2,o2} %Connects the outputs with a vertex.
  \fmf{photon,label=$\gamma(q)$}{v1,v2}
 \fmfblob{.16w}{v1}
 \fmfdot{v2}
%labels the conneting line.
\end{fmfgraph*}
 }
\end{fmffile}
\end{tabular}
\\
\caption{One photon exchange for $\bar{p}\,+\,p \to l^-\,+\,l^+$
in the $s$ channel.} \label{fey3}
 \end{center}
\end{figure}
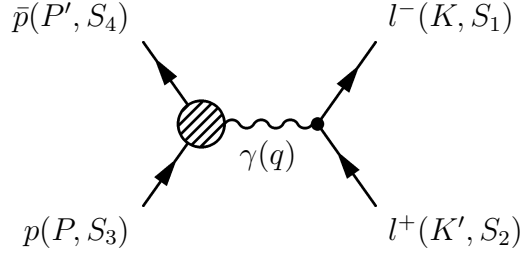
For the reaction $\bar{p}\,p \to l^+\,l^-$
 we will be using the following momentum four
 vectors in the centre of mass system:
 \begin{eqnarray}
\label{mom}
  P^\mu &=& \left(\,E,\,0,\,0,\,p\right)\nonumber
\\
 P'^\mu &=&  \left(\,E,\,0,\,0,\,-p\right)\nonumber
 \\
 K^\mu &=& \left(\,E,\,k\,\sin \theta,\,0,\,k\,\cos
 \theta\right)\nonumber
  \\
  K'^\mu &=& \left(\,E,\,-k\,\sin \theta,\,0,\,-k\,\cos \theta\right)
 \end{eqnarray}
 where the centre of mass scattering angle $\theta$ is given by
 \begin{eqnarray}
  \cos\theta &=& \frac{t - u}{\tstrut
 \sqrt{s - 4\,m^2}\,\sqrt{s - 4\,M^2}}
 \end{eqnarray}
 In order to include the spin-up and spin-down vectors
in the following calculation we introduce the
 variables $\epsilon_1,\,\epsilon_2,\,\epsilon_3,\,
\epsilon_4 $ which are either $\pm1$ depending on whether
 the spin is up or down.
 \begin{eqnarray}
 \label{spin}
 S_1 &=& \frac{\epsilon_3}{m}\,(k,\,E\,\sin \theta,\,0,\,E\,\cos
 \theta)\nonumber
 \\\nonumber
  \\\nonumber
 S_2 &=& \frac{\epsilon_4}{m}\,(k,\,-E\,\sin \theta,\,0,
 \,-E\,\cos \theta) \nonumber
 \\
 \\ \nonumber
 S_3 &=& \frac{\epsilon_1}{M}\,(p,\,0,\,0,\,E) \nonumber
 \\\nonumber
  \\ \nonumber
 S_4 &=& \frac{\epsilon_2}{M}\,(p,\,0,\,0,\,-E)\nonumber\,.
\end{eqnarray}
 These spin four vectors are normalised to $-1$
and are perpendicular to the corresponding
 momentum four vector, e.g. $ P_{\mu}\,S_{1}^{\mu} = 0 $.
 The momentum of the proton and the lepton are $p$ and $k$
 respectively and are related to the energy, $E= \sqrt{s}/2$, as
 \begin{eqnarray}
  p &=& \frac{1}{2}\sqrt{s - 4\,M^2}\nonumber
 \\
 k &=& \frac{1}{2}\sqrt{s - 4\,m^2}\,.\nonumber
\end{eqnarray}

\end{document}